# GRIDDED RF GUN DESIGN FOR SRF LINAC APPLICATIONS

I. Gonin†, C. Edwards, S. Kazakov, T. Khabiboulline, T. Nicol, A. Saini, N. Solyak,
J.C.T. Thangaraj, V. Yakovlev, FNAL, Batavia, USA
M. Curtis, K. Gunther, HeatWave Labs, Inc., Watsonville, CA, USA

## Abstract

The thermionic gridded gun is described which generates short electron bunches for further acceleration in a Nb3Sn conduction cooled SRF linac. The gun is built into the first cavity of the 250 keV injector [1] for the 20 kW, 10 MeV, 1.3 GHz CW conduction cooled one-cavity linac. The beam current is 2 mA. The RF gun design is presented as well as the results of perveance measurements, which are in a good agreement with the design parameters. The design of the RF resonator of the gun is presented also. The beam generated by the gun is matched to the injector to provide lack of current interception in the SRF cavity.

## GENERAL

The concept of compact linear electron accelerators for industrial application suggested in [2] is based on use of Nb3Sn conduction cooled SRF cavities. This concept allows both high energy ≳ 10 MeV and high power of MW level. One of the most serious limitations of the beam power is the beam current interception in the SRF cavity cooled by few cryo-coolers which provide power removal of ~2-2.5 W each at 4.5 K. Therefore, power loss in the SRF cavity caused by intercepted electrons should not exceed ~1 W, or ~ 1.e-6. It means that a source of electrons should provide short bunches without tails, which may be defocused by RF field and reach the cavity surface. In addition, the beam should have considerably low transverse emittance preventing strong bam diameter increase in the cavity. To generate these short bunches, we suggested to use gridded RF gun [3]. RF field is applied to the gun cathode – grid gap together with the bias deceleration DC voltage. This scheme provides electron injection in very narrow phase domain, few degrees, and therefore, generation of very short bunches. On the other hand, this scheme provides lack of long bunch tails. Initially the gun has been designed for 650 MHz, 1.8 MeV linac prototype which is under construction at IARC [3]. It will be installed directly into the SRF cavity. For 1.3 GHz prototype [2] a special room-temperature injector has been designed [1] which provides the 250-300 keV beam for the SRF accelerator. This concept potentially allows higher beam power than the scheme with immersed gun, because high power and consequently high beam current (up to 100 mA at very short, <10° pulses) requires considerably big cathode diameter, and therefore, large black body radiation and cavity contamination caused by the cathode material evaporation. The design of a thermionic electron source which can either be directly connected to a superconducting cavity or be part of a normal conducted injector cavity is described.

The direct connection option is applied in a prototype 1½ cell 650 MHz SRF cavity capable of delivering a 12.5 mA average beam current with a beam power of 20 kW. As an external option we present the development of a CW normal conducting 1.3 GHz RF injector which consists of a gridded RF gun integrated with the first cell of a copper booster cavity. The electron source concept is presented including the cathode-grid assembly and the gun resonator design.

## THE RF GUN DESIGN

The RF gun consists of the cathode-grid assembly and the gun RF resonator. Figure 1 show a 3D cross-section of the cathode-grid assembly [3]. The assembly contains the cathode-grid part, the transition support part, the socket part with the ceramic window which subdivide the vacuum cathode part and air RF resonator, and the socket.

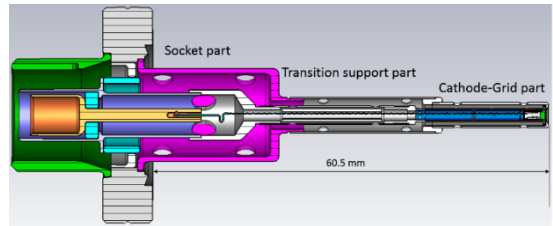

Figure 1: Cross-section of the cathode-grid assembly.

The beam optics is determined by the shape of the grid extension electrode with Pierce angle. It has an impregnated cathode of 612M type with a 1.8 eV work function to provide electron current density greater than 10 A/cm$^2$ at emitter temperature of 1050ºC. The cathode Ø = 1.5 mm to provide low heat load (< 1W). The grid is placed at the distance of 150 microns from the cathode. The cathode has planar shape. The grid is attached to the Pierce angle electrode. The grid transparency is higher than 80%. Planar geometry of the cathode and grid is less sensitive to dimensional errors and misalignments. A heater filament should be designed to avoid the heater magnetic field on the cathode. The heater power level < 2 W. The gun cathode-grid unit engineering design is made by HeatWave Labs (HWL).

Figure 2 shows the cathode-grid area with details of the heater. An outer conductor and the grid are grounded. The cathode and heater are at a DC bias voltage. To minimize the heater power, the cathode is mounted using a support sleeve made of 25 µm (0.001") thick molybdenum/rhenium alloy with an additional hole pattern to reduce the thermal path to emulate a 12.5 µm thick sleeve. To reduce thermal radiation, a special thermal shield of a 25 µm thick Mo/Re has been added to the design.

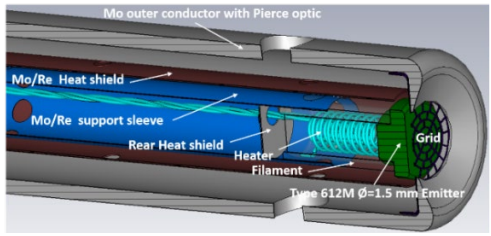

Figure 2: Detailed view of the cathode-grid area.

Two similar cathode-grid units have been manufactured by HWL, SN 10746 and SN 10747, see Fig. 3. The length of the assembly is 186 mm.

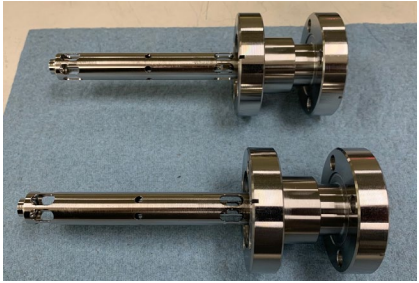

Figure 3: The cathode-grid assemblies.

To characterize the units, the gun perveance has been measured. The results of the perveance measurement are show in Fig. 4 for both units. The perveance has been compared to MICHELLE [4] simulations, that allows to determine the cathode-grid gap. For SN 10746 the gap is 150 microns, and SN 10747 it is 175 microns, which is in excellent agreement with the design.

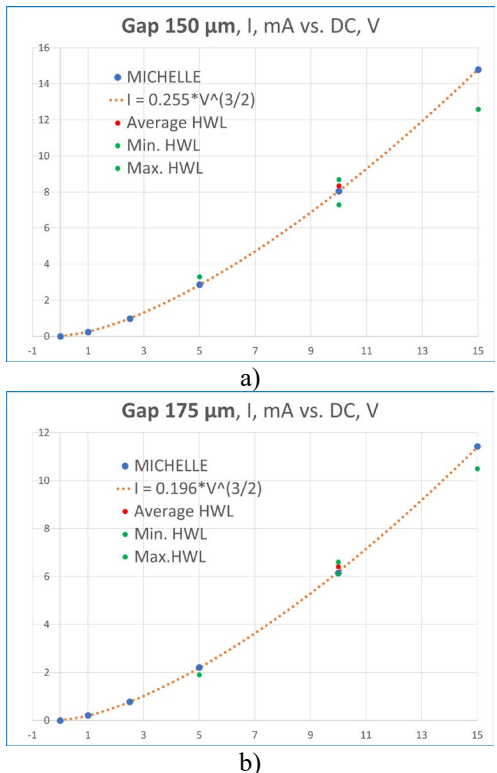

Figure 4: The gun perveance measurements for SN 10746 (a) and SN 10747 (b).

The gun cathode-grid unit is connected to the RF resonator. Electro-magnetic and mechanical design for the 1.3 GHz RF gun resonator is completed and shown in Fig. 5. The RF gun parameters are shown in Table 1. The resonator has the input coupler, the frequency tuning insert, heater and bias connectors and air-cooling inlet, which may be necessary for future operation at higher current and RF voltage. Note that the temperature rise caused by RF heating the gun is small, because in the latest design the socket par and transition support part are made of copper except the sleeves next to ceramics made of Kovar. The gun RF resonator excitation by the field of the first injector cavity penetrating the gun gap through the grid has been analysed, it is negligible.

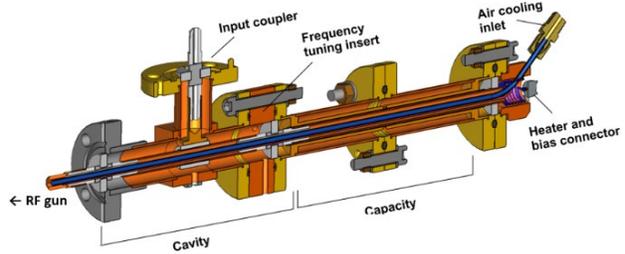

Figure 5: The gun RF resonator having the input coupler, frequency tuning insert and connectors.

Table 1: The RF gun parameters.

| Parameter | Value | Unit |
| --- | --- | --- |
| DC voltage | 30 | V |
| RF voltage | 65 | V |
| Beam current | 2 | mA |
| R/Q | 12 | Ohm |
| $Q_0$ | 1004 | |
| Shunt impedance | 12 | kOhm |
| Bandwidth | 1.3 | MHz |
| Wall power dissipation | 0.34 | W |
| Input power | <1 | W |
| Losses on the grid | 40 | mW |
| Grid transparency | 80 | % |
| Temperature rise in the gun caused by RF | 34 | °C |

The gun is installed in the 1st cavity of the 1.3 GHz injector (Figure 6), the gun beam dynamics is matched to the beam dynamics in the injector providing lack of current interception in the SRF cavity installed after the injector [1,5], see Figure 7.

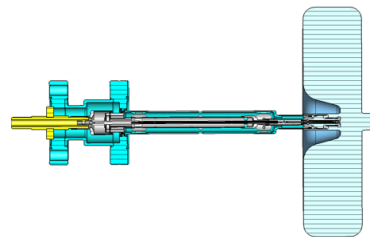

Figure 6: The RF gun integrated to the injector cavity.

Note that for future 200 kW linac the beam current should be of 20 mA. In present gun the main current limitation is the grid heating by the elections. This limitation can be mitigated using the shadow cathode, which is under development at HWL, see Figure 8.

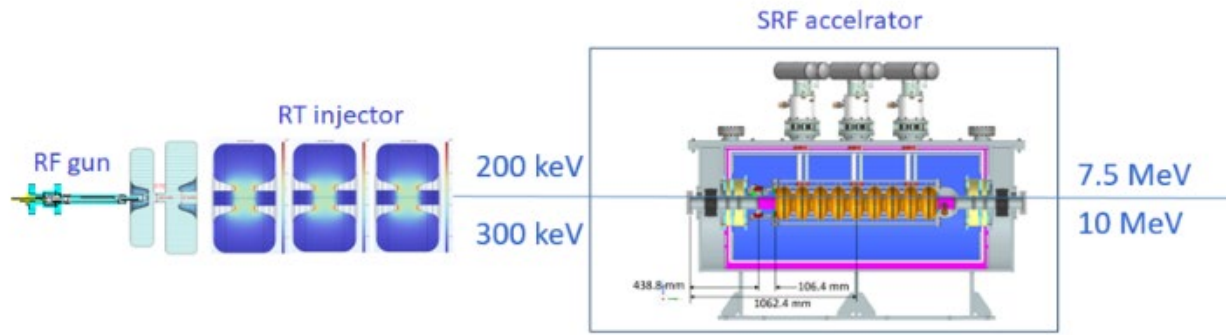

Figure 7: The SRF 1.3 GHz, 10 MeV, 20 kW linac layout.

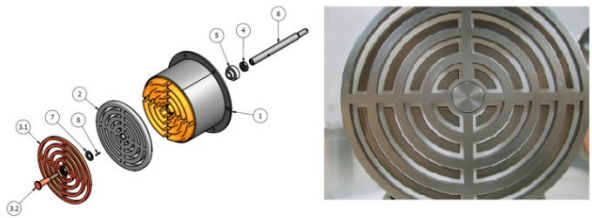

Figure 8: Shadow cathode under development by HWL.

Shadow cathode will allow the beam current of 25 mA.

## CONCLUSION

RF gun provides short bunches without tails necessary to avoid beam current interception in the conduction cooled SRF cavity. The gun design is performed. The gun is manufactured and tested showing design parameters. The gun optics is matched to the injector and the SRF linac to provide required output beam parameters and lack of current intercepting in the SRF cavity. The RF resonator design is ready including the input coupler, DC input, windows, cooling. Further development of the RF gun design (shadow cathode) is in progress. It will allow the average current up to 25 mA (limitation is caused by grid heating) necessary for 10 MeV, 200 kW linac.

## ACKNOWLEDGEMENTS

The experiments described and the resulting data presented herein, unless otherwise noted, were funded under PE 0603119A, Project B03 "Accelerator Technology for Ground Maneuver", managed by the US Army Engineer Research and Development Center. The work described in this presentation was conducted at Fermilab. Permission was granted by Fermilab and ERDC to publish this information.